\documentclass[12pt]{article}

\usepackage{amsmath,amssymb,amsthm,epsf,epsfig}
\usepackage{amsmath,amssymb}
\begin{document}

\begin{center}
{\bf {\Large Planck Scale Effect in   the Entropic Force Law }}\\
\vspace{2cm}

 Subir Ghosh
  \vspace{1ex}

Physics and Applied Mathematics Unit, \\
Indian
Statistical Institute, \\
203 B.T.Road, Kolkata 700108, India. \end{center}

\vspace{3ex}

\begin{abstract}
In this note we generalize the quantum uncertainty relation
proposed by Vancea and Santos [7] in the entropic force law, by
introducing Planck scale modifications. The latter is induced by
the Generalized Uncertainty Principle. We show that the proposed
uncertainty relation of [7], involving the entropic force and the square of particle position,
gets modified from the consideration of a  minimum measurable length, (which can be the Planck length).
\end{abstract}

Very recently Verlinde \cite{ver} has conjectured that the origin
of Newtonian gravity and (second) law of dynamics might be
entropic in nature. This means that the conventional forces can
originate from maximization of entropy principle as in
 thermodynamics of macroscopic systems. This framework attempts to
 establish thermodynamics as the fundamental principle. These
 ideas are further strengthened by the deep connection between
 thermodynamics and black hole physics, as advocated by previous
 workers \cite{be,un,ja,pad}. A number of subsequent developments
 \cite{others}
 in various directions have been reported after Verlinde's work.

 Deep physical insight coupled with very simple algebra has led \cite{ver} from the entropy principle to
 the second law of Newtonian dynamics,
 \begin{equation}
P=ma
 \label{pmf}
 \end{equation}
where the force  $P$ on a particle is related to its mass $m$ and
acceleration  $a$. This is derived from the the first law of
thermodynamics,
 \begin{equation}
\Delta W=T\Delta S=F\Delta x
 \label{2t}
 \end{equation}
where the variation in energy $W$ of a macroscopic system is
expressed in terms of its equilibrium temperature $T$ and change
in entropy $S$. Furthermore, as is customary in thermodynamics,
this can also be identified as the work done by a generalized
force $F$ for the displacement $\Delta x$. $F$ is termed as the entropic force. The essential cog in
this analysis is the conjecture that there is a variation in
entropy associated with the holographic screen, (that separates
the emerged spacetime from the sector to be emerged), as a
particle of mass $m$ approaches very close (of the order of the
particle Compton length $l_c=\hbar /mc$) to the screen. Explicitly
it is postulated that \cite{ver}
 \begin{equation}
\Delta S=2\pi k_B\frac{\Delta x}{l_c},
 \label{s}
 \end{equation}
where $\Delta x$ is the distance of the particle from the screen.
Furthermore, Verlinde \cite{ver} identifies the thermodynamic
temperature $T$ in (\ref{2t}) to the Unruh temperature \cite{un}
$T_U$
 \begin{equation}
T_U=\frac{\hbar a}{2\pi k_Bc},
 \label{un}
 \end{equation}
where $a$ denotes the acceleration of the observer who experiences
$T_U$. Combining all these a simple algebra leads to Newton's law
(\ref{pmf}).

In a recent paper Vancea and Santos \cite{vs} have opened another
line of thought. They have pointed out \cite{vs} that the postulate
(\ref{s}) as well as the expression for  Unruh temperature
(\ref{un}) are essentially quantum in nature, with the explicit
presence of $\hbar$, although $\hbar$ does not show up in the
final outcome, Newton's classical law of motion. Indeed, as far as the latter is concerned, this is
as it should be but one can expect quantum corrections to Newton's
second law of motion and the law of gravitation. It has been
emphasized by Padmanabhan already \cite{pad} that gravity is
intrinsically quantum in the holographic approach as the
"classical limit" $\hbar \rightarrow 0$ leads to a divergence in
the Newton constant. However, more importantly for our purpose, is
the introduction of Planck length $L_P$ in \cite{pad} as a
fundamental scale for counting the number of micro cells in the
holographic screen that yields the entropy. In \cite{vs} the
possibility of quantum correction to Newton's law of motion (and
gravitation) has been speculated. In the present work we have
generalized the work in \cite{vs} by incorporating the other
essential length scale $L_P$ in the generalized Newton's law.

 More explicitly, in \cite{vs} the authors have considered the possibility of
having an uncertainty in the entropy in the holographic screen
originating from the fact that there are inherent {\it{quantum}} uncertainties in the
position and momentum of a particle. This comes primarily because
the information (or entropy) associated with the screen depends
linearly on the distance of the test particle from the screen. Keeping this in mind, in \cite{vs}, the
relation (\ref{s}) has been generalized to
 \begin{equation}
\delta S=2\pi k_B(\frac{\delta x}{l_c}+\frac{\delta p}{mc}).
 \label{vvs}
 \end{equation}
In fact another alternate form of $\delta S$ has also been
suggested in \cite{vs} but we choose here the minimal form. One
can think of the denominators as $l_c=\hbar/(mc)$ (for $\delta x$)
and $\hbar /l_c=mc$ (for $\delta p$). In \cite{vs} the variations
$\delta x$ and $\delta p$ are to considered as quantum
uncertainties, obeying the Heisenberg uncertainty relation
 \begin{equation}
\delta x\delta p \ge \frac{\hbar}{2}.
 \label{q}
 \end{equation}
Indeed in the classical case $\delta x$ reduces to the separation
$\Delta x$ between the screen and the particle and $\delta p =0$
as the Heisenberg uncertainty relation does not apply. Then (\ref{vvs}) reduces to the Verlinde formula (\ref{s}). Otherwise, one
can replace $\delta p$ in (\ref{vvs}) by $\delta p=\hbar /(2\delta
x)$ to obtain a quantum corrected Newton's second law of motion
\cite{vs}
\begin{equation}
F(\delta) = ma + \frac{\hbar}{2m} \left( \frac{\hbar a}{c^2} - p
\right) (\delta x)^{-2}. \label{qf1}
\end{equation}
The above yields an uncertainty relation, first proposed in
\cite{vs},
\begin{equation}
\delta F (\delta x)^2 \geq \frac{\hbar}{2m} \left( \frac{\hbar
a}{c^2} - p \right). \label{qf2}
\end{equation}
Furthermore, using the classical (Newtonian) expression for
gravitational acceleration $a=(GM)/R^2$ a quantum corrected
Newton's law for gravitation is also obtained \cite{vs}, to first
order in $\hbar$,
\begin{equation}
F = G\frac{Mm}{R^2} + \frac{\hbar }{2m} \left( G \frac{\hbar
M}{R^2 c^2} - p \right) (\delta x)^{-2}. \label{qg}
\end{equation}

In this perspective, once the quantum correction has been
introduced, it is indeed natural to consider the Generalized
Uncertainty Principle (GUP) \cite{gupaper} instead of the
Heisenberg relation (\ref{q}). In recent years there has been a
lot of activity, studying the consequences of GUP (in
non-relativistic \cite{gup1} and cosmological \cite{gupc}
contexts) since it introduces a minimum length scale, generally
considered to be the Planck length $\l_{P}^2=G\hbar/c^3$. Implications of GUP in the context of Information Theory and Holography, that are very relevant for the present work, was studied in \cite{kempf1}.  A
non-trivial prediction of generic quantum gravity theories (such
as String Theory) and black hole physics is the existence of a
minimum measurable length \cite{gupaper}. Heuristically, a minimum length cut off
is needed to avoid the paradox that localization of an event below
Planck length can generate sufficient energy density to create a
black hole, thus rendering the event itself unobservable. For the
more numerically minded, the Schwarzschild radius near the Planck
scale, $l_s\sim (M_pG)/c^2\sim \sqrt{(Gh)/c^3}$ becomes comparable
to the Compton wavelength, $l_c\sim h/(M_pc)\sim \sqrt{(Gh)/c^3}$.
Incidentally both are of the order of the Planck length. This
cherished length scale can be induced by the GUP, of the following
form first provided in \cite{kempf},
\begin{equation}
\delta x_i \delta p_i  \geq  \frac{\hbar}{2} [ 1 + \beta ((\delta
p)^2 + <p>^2)+
 2\beta ( \delta p_j^2 + <p_j>^2) ]~,~i=1,2,3
 \label{gun}
 \end{equation}
where $p^2 = \sum\limits_{j=1}^{3}p_{j}p_{j}$,
$\beta \sim 1/(M_{P}c)^2=\ell_{P}^2/2\hbar^2$, $M_{P}=$ Planck mass,
and $M_{P} c^2=$ Planck energy $\approx 1.2 \times 10^{19}~GeV$.
Subsequently, the  Hilbert space representation of the generalized uncertainty relation was developed in \cite{kmm}. This is indeed crucial for formulating the quantum mechanics in a phase space with a non-commutative  structure that is  compatible with the above GUP.

Our main result, in this short note, is to show that the minimum
length scale, via the GUP, modifies the
new force-position uncertainty relation (\ref{qf2}), suggested in
\cite{vs}. In particular we find,
\begin{equation}
\delta F (\delta x)^2 \geq \nu\frac{\hbar}{2m} \left( \frac{\hbar
a}{c^2} - p \right),
 \label{fx}
 \end{equation}
where $\nu =1+\frac{\hbar^2\beta}{4\delta x^2}$ is the GUP induced $\beta $-dependent correction. Taking the
minimum value of $\delta x$ to be $l_P$ we find the maximum value of $\nu $ to be $\nu \sim 1+5/4$. (The
exact numerical factor should not be taken too seriously.) We will
derive this relation in rest of the note.

We start by simplifying the GUP to one dimension,
 \begin{equation}
\delta x\delta p\geq \frac{\hbar }{2}(1+\beta (\delta p) ^2).
 \label{gup}
 \end{equation}
Solving the saturation condition to $O(\beta )$ we find
\begin{equation}
\delta p=\frac{\hbar}{2\delta x}(1+\frac{\hbar^2\beta}{4}).
 \label{pvar}
 \end{equation}
and demanding the reality of $\delta p$  yields the inequality,
 \begin{equation}
\delta x\geq \sqrt \beta \hbar,
 \label{xg}
 \end{equation}
with $\delta x= \sqrt \beta \hbar $ being the minimum measurable
length. This is taken as Planck length $L_P$. Thus, following the same
steps as in \cite{vs} we obtain the expression for entropic force
consistent with GUP (\ref{gup}),
 \begin{equation}
F=ma+\frac{\hbar}{2mc^2(\delta x)^2}(a\hbar
-pc^2)(1+\frac{\hbar^2\beta}{4(\delta x)^2}).
 \label{fg}
 \end{equation}
As before \cite{vs}, defining the uncertainty in $F$ to be $\delta
F=F-ma$ \cite{vs} we obtain
 \begin{equation}
\delta F(\delta x)^2\geq \frac{\hbar}{2mc^2}(a\hbar
-pc^2)(1+\frac{\hbar^2\beta}{4(\delta x)^2}) \equiv \nu\frac{\hbar}{2mc^2}(a\hbar
-pc^2),
 \label{fg1}
 \end{equation}
in conjunction with the minimum measurable length $\delta
x\geq \sqrt \beta \hbar \sim L_P$.
 This is the major result of our paper. The effect of the $\beta $-correction can be seen from the relation,
\begin{equation}
 (\delta x)^2_\beta =\frac{1}{2}[(\delta x^2)+(\delta x)^2\sqrt{1+\frac{\hbar ^2\beta}{(\delta x)^2}}~],
\end{equation}
where we considered the saturation values of $\delta x^2$ for the two cases, ( $(\delta x)^2$ for $\beta =0$ \cite{vs}
and $(\delta x)^2_\beta$ for
$\beta \ne 0$), for the same $\delta F$.

Incidentally this ties up nicely with the idea of Padmanabhan
\cite{pad} that a naive $\hbar \rightarrow 0$ limit is not
acceptable as it leads to a diverging Newton's constant. On the
other hand the new uncertainty relation (\ref{qf2}), proposed in
\cite{vs}, has the possibility of generating large uncertainty
$\delta F\rightarrow \infty $ as $\delta x \rightarrow 0$.
However, our analysis shows that the minimum length scale $L_P$
once comes again to the rescue by weakening the undertainty relation of \cite{vs} thus
 allowing finite values of  $\delta F$.

Let us conclude by pointing out the significance of our analysis
in the context of experimental/observational validity to ascertain
the consistency of the entropic scenario, its quantum correction,
as well as the hypothesis of minimum length scale. For this, two
features of our results should be worth mentioning: (i) The
correction factor $\nu$ in (\ref{fg1}) is not damped by the Planck
scale. (ii) The purported quantum corrections appear in the
classical laws of motion. Hence it is conceivable that one might
conduct suitable experiments where additive physical quantities
such as mass, energy etc. appear as observables and their
cumulative effects might leave a substantial signature of quantum
or even the Planckian regime in a classical scenario.
\vskip .5cm
{\it{Acknowledgment}}: It is a pleasure to thank Professor I. V. Vancea for correspondence. I am also grateful to the Referee for comments.

\end{document}